\documentclass[12pt,a4paper,final]{iopart}

\usepackage{iopams}
\usepackage{graphicx}
\usepackage[symbol]{footmisc}
\usepackage[breaklinks=true,colorlinks=true,linkcolor=blue,urlcolor=blue,citecolor=blue]{hyperref}

\begin{document}

\title[Ground state cooling in a hybrid optomechanical system]{Ground state cooling in a hybrid optomechanical system with a three-level atomic ensemble}

\author{
Tan Li$^{1,2}$\footnote[1]{\label{fn:note1}These authors contributed equally to this work.},
Shuo Zhang$^{1,2}$\ref{fn:note1},
He-Liang Huang$^{1,2}$,
Feng-Guang Li$^{1,2}$,
Xiang-Qun Fu$^{1,2}$,
Xiang Wang$^{1,2}$ and
Wan-Su Bao$^{1,2}$\footnote[2]{Author to whom any correspondence should be addressed.}}

\address{$^1$ Henan Key Laboratory of Quantum Information and Cryptography, Zhengzhou Information Science and Technology Institute, Zhengzhou, Henan 450001, China}
\address{$^2$ Synergetic Innovation Center of Quantum Information and Quantum Physics, University of Science and Technology of China, Hefei, Anhui 230026, China}

\ead{2010thzz@sina.com}

\begin{abstract}
Cooling mechanical resonators is of great importance for both fundamental study and applied science.
We investigate the hybrid optomechanical cooling with a three-level atomic ensemble fixed in a strong excited optical cavity.
By using the quantum noise approach, we find the upper bound of the noise spectrum and further present three optimal parameter conditions, which can yield a small heating coefficient, a large cooling coefficient, and thus a small final phonon number.
Moreover, through the covariance matrix approach, results of numerical simulation are obtained, which are consistent with the theoretical expectations. It is demonstrated that our scheme can achieve ground state cooling in the highly unresolved sideband regime, within the current experimental technologies. Compared with the previous cooling methods, in our scheme, there are fewer constraints on the drive strength
of atomic ensemble and number of atoms in the ensemble. In addition, the tolerable ranges of parameters for ground state cooling are extended. As a result, our scheme is very suitable for experiments and can be a guideline for the research of hybrid optomechanical cooling.
\end{abstract}

\vspace{2pc}

\submitto{\JPB}
\maketitle

\section{Introduction}

Cooling of the mechanical resonator (MR) has attracted considerable research attention, which is a crucial step for the applications of MR, such as quantum transducers \cite{Stannigel2010, Stannigel2011, Bochmann2013, Andrews2014}, quantum computing \cite{Rips2013, Huang2017}, precision metrology \cite{Zhang2012, Mahajan2013}, macroscopic quantum physics \cite{Genes2009, Pikovski2012}, and so on.
For this reason, people have proposed various MR cooling methods.
Among them, the optomechanical cooling \cite{Kippenberg2008,Marquardt2009,Liu2013a,Liu2013,Aspelmeyer2014} is a kind of significant direction, which makes the MR couple to an optical cavity through the radiation force.

In this direction, the typical one is the sideband cooling method \cite{Marquardt2007,Wilson-Rae2007}, where the drive laser of the cavity is red detuned by a mechanical frequency $\omega_{m}$, analogous to the cooling of an ion or atom \cite{Wineland1979,Zippilli2005, Zhang2014a, Zhang2014}.
Thus, the Anti-Stokes sideband is enhanced due to the cavity resonance, while the off-resonant Stokes sideband is greatly suppressed in the resolved sideband regime, and the MR can be cooled to the ground state \cite{Chan2011, Teufel2011}.
However, for some systems the resolved sideband condition (that is, the cavity decay $\kappa$ is far less than $\omega_{m}$) is not easy to fulfill.
Then, the suppression of Stokes sideband could not be sufficient, which prevents the ground state to be reached.
For this, many hybrid optomechanical cooling methods have been proposed, which hybridize the optomechanical system with another assisted system.
For example, an atomic ensemble \cite{Genes2009a,Genes2011,Ge2013,Vogell2013,Joeckel2014,Bennett2014,Nie2015,Chen2015,Zeng2015}, an atom \cite{Breyer2012,Yi2014,Zhang2014b}, a MR \cite{Ojanen2014,Dong2015,Liu2015a}, an optical cavity \cite{Li2011,Gu2013,Guo2014,Liu2015,Kim2017,Feng2017}, or a subsystem composed of an optical cavity and another system \cite{Han2014,Bariani2014,Degenfeld-Schonburg2016}.

The pioneering work \cite{Genes2011} presented a hybrid optomechanical cooling method with a three-level atomic ensemble.
However, the collective atom-cavity coupling strength $g_N$ is assumed to be much larger than the drive strength $\Omega_r$ of atomic ensemble, which consequently leads to a constraint condition $N\gg\left|\bar{a}\right|^{2}$, where $N$ is the number of atoms and $\left|\bar{a}\right|^{2}$ is the number of intracavity steady-state photons. Due to the cooling rate scales with number of intracavity photons, a given $N$ will restrict the maximum achievable cooling rate; meanwhile, achieving a larger cooling rate requires a larger $N$, which brings extra limitations for the experimental implementation. Moreover, the theoretical analysis only takes into account the case that drive of the cavity is red detuning by $\omega_{m}$, whether or not other cases of detuning is allowed for ground state cooling remains unknown. In addition, some system parameters may not be chosen as optimal, for example the atomic detunings, which may weaken the cooling performance and can be found as well in other cooling schemes \cite{Liu2015a,Bariani2014}.

In this paper, we expect to remove these limitations (that is, $g_N\gg\Omega_r$ and $N\gg\left|\bar{a}\right|^{2}$) in this hybrid optomechanical system by using quantum noise approach, and give the optimal parameter cooling conditions to enhance the cooling coefficient and suppress the heating coefficient in a more direct way, which is expected to provide a better guideline for experiments.

The paper is organized as follows. In Sec.~\ref{sec:System_Decription}, we describe the system using Hamiltonian, Langevin equations and master equation.
In Sec.~\ref{sec:Analytical_Results}, by using the quantum noise approach, we calculate the noise spectrum, heating/cooling coefficient and final phonon number, analyze the upper bound of noise spectrum and further give the optimal parameter conditions, including the atomic and cavity detunings.
In Sec.~\ref{sec:Numerical_Results}, we numerically simulate the time evolution of mean phonon number, and systematically explore the dependence of final phonon number on the system parameters, which demonstrates that ground state cooling is achievable in the highly unresolved sideband regime.
In Sec.~\ref{sec:Discussion}, we compare the existing related works and this work, followed by a brief conclusion in Sec.~\ref{sec:Conclusion}.

\section{Model \label{sec:System_Decription}}

\begin{figure}[b!]
\centering
\includegraphics[width=8.0cm]{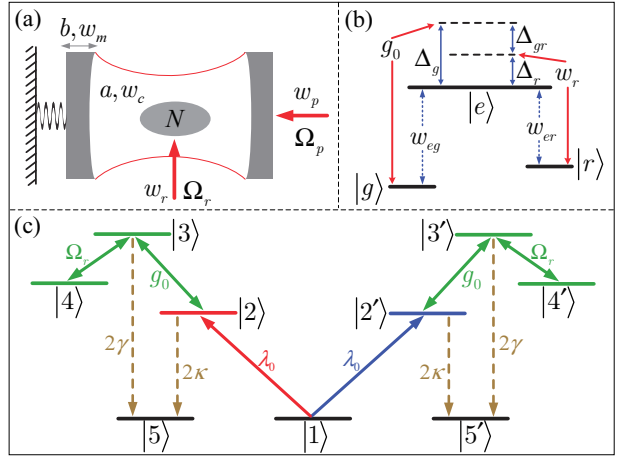}
\protect\caption{(a) The schematic of the hybrid optomechanical cooling model. An three-level atomic ensemble is fixed in the FP cavity.
(b) The energy-level diagram of atom in the ensemble. $\left|e\right\rangle$ is the excited state, $\left|g\right\rangle$ and $\left|r\right\rangle$ are the ground states.
(c) The illustration of transitions among system states.
$\left|1\right\rangle=\left|g,m,n\right\rangle$,
$\left|2\right\rangle=\left|g,m+1,n-1\right\rangle$,
$\left|2'\right\rangle=\left|g,m+1,n+1\right\rangle$,
$\left|3\right\rangle=\left|e,m,n-1\right\rangle$,
$\left|3'\right\rangle=\left|e,m,n+1\right\rangle$,
$\left|4\right\rangle=\left|r,m,n-1\right\rangle$,
$\left|4'\right\rangle=\left|r,m,n+1\right\rangle$,
$\left|5\right\rangle=\left|g,m,n-1\right\rangle$,
$\left|5'\right\rangle=\left|g,m,n+1\right\rangle$,
where $\left|\alpha,m,n\right\rangle$
($\left|\alpha\right\rangle$ = $\left|g\right\rangle$, $\left|r\right\rangle$ or $\left|e\right\rangle$)
represents a state of system with
each atom at $\left|\alpha\right\rangle$, $m$ photons in mode $a$, $n$ phonons in mode $b$.
\label{fig:model}}
\end{figure}

As is presented in figure~\ref{fig:model}(a), the cooling scheme is composed of an atomic ensemble and an optical Fabry-P\'{e}rot (FP) cavity  with a movable end mirror.
The atomic ensemble consists of $N$ identical three-level atoms, and is assumed to be fixed in the cavity without mechanical oscillation.
The FP cavity (atomic ensemble) is driven by a pump laser with frequency $\omega_p$ ($\omega_r$) and strength $\Omega_p$ ($\Omega_r$).
The optical (mechanical) mode we concerned is denoted as $a$ ($b$) with frequency $\omega_c$ ($\omega_m$).

As shown in figure~\ref{fig:model}(b), the atom in the ensemble is in the $\Lambda$-type configuration.
The single excited state is denoted as $\left|e\right\rangle$, while the two ground states are denoted as $\left|g\right\rangle$ and $\left|r\right\rangle$.
$\omega_{eg}$ ($\omega_{er}$) represents the energy-level frequency separation between
$\left|e\right\rangle$ and $\left|g\right\rangle$ ($\left|r\right\rangle$).
The transitions $\left|g\right\rangle \leftrightarrow\left|e\right\rangle$ and $\left|r\right\rangle\leftrightarrow\left|e\right\rangle$ interact with the cavity field and drive laser of ensemble, respectively, where $\Delta_g=\omega_p-\omega_{eg}$ and $\Delta_r=\omega_r-\omega_{er}$ are the corresponding detunings, with a difference $\Delta_{gr}=\Delta_g-\Delta_r$.

The state of the whole system can be represented as $\left|\alpha,m,n\right\rangle$, where $\left|\alpha\right\rangle$ ($= \left|g\right\rangle$, $\left|r\right\rangle$ or $\left|e\right\rangle$) describes the state of atom, $m$ is the photon number, and $n$ is the phonon number.
In figure~\ref{fig:model}(c), we present the transitions among different system states, where $\left|1\right\rangle \leftrightarrow\left|2\right\rangle$ ($\left|1\right\rangle\leftrightarrow\left|2'\right\rangle$), $\left|2\right\rangle \leftrightarrow\left|3\right\rangle$ ($\left|2'\right\rangle\leftrightarrow\left|3'\right\rangle$), $\left|3\right\rangle \leftrightarrow\left|4\right\rangle$ ($\left|3'\right\rangle\leftrightarrow\left|4'\right\rangle$) indicate the interaction between MR and optical cavity, optical cavity and atomic ensemble, atomic ensemble and its drive laser, with coupling rates $\lambda_0$, $g_0$ and $\Omega_r$, respectively.
$\left|2\right\rangle \leftrightarrow\left|5\right\rangle$ ($\left|2'\right\rangle\leftrightarrow\left|5'\right\rangle$) and $\left|3\right\rangle \leftrightarrow\left|5\right\rangle$ ($\left|3'\right\rangle\leftrightarrow\left|5'\right\rangle$) indicate the optical decay of FP cavity and the atomic spontaneous emission from $\left|e\right\rangle$ to $\left|g\right\rangle$.
Note that, here the emission from $\left|r\right\rangle$ to $\left|g\right\rangle$ is assumed to be negligible weak, which is reasonable for three-level atoms in usual.
Under suitable parameter conditions, assisted by the atomic ensemble, the heating process $\left|1\right\rangle \leftrightarrow\left|2'\right\rangle$ can be eliminated by the destructive interference, while the cooling process $\left|1\right\rangle \leftrightarrow\left|2\right\rangle$ can be enhanced or remain unchanged, which is expected to yield a better cooling result.

\subsection{Hamiltonian}

The system Hamiltonian reads (with $\hbar=1$) \cite{Scully1997}
\begin{equation}
H=H_{0}+H_{I}+H_{pump},
\label{eq:Origianl_Hamiltonian}
\end{equation}
where
\begin{eqnarray}
H_{0} & = & \omega_{c}a^{\dagger}a+\omega_{m}b^{\dagger}b+\omega_{eg}\Sigma_{j=1}^{N}\sigma_{ee}^{j}
+\left(\omega_{eg}-\omega_{er}\right)\Sigma_{j=1}^{N}\sigma_{rr}^{j},\nonumber \\
H_{I} & = & \lambda_{0}a^{\dagger}a\left(b^{\dagger}+b\right)
+g_{0}\Sigma_{j=1}^{N}\left(a\sigma_{eg}^{j}+a^{\dagger}\sigma_{ge}^{j}\right),\nonumber \\
H_{pump} & = & \Omega_{p}\left(a^{\dagger}e^{-i\omega_{p}t}+ae^{i\omega_{p}t}\right)
+\Omega_{r}\Sigma_{j=1}^{N}\left(e^{-i\omega_{r}t}\sigma_{er}^{j}+e^{i\omega_{r}t}\sigma_{re}^{j}\right).
\label{eq:three_parts_of_Origianl_Hamiltonian}
\end{eqnarray}
The first part $H_{0}$ is the free Hamiltonian of the optical cavity, MR, and atomic ensemble, where $\sigma_{\alpha\beta}^{j}=\left|\alpha\right\rangle \left\langle \beta\right|$ ($1\le j\le N$) is the transition operator of $j-{\rm{th}}$ atom from $\left|\beta\right\rangle $ state to $\left|\alpha\right\rangle $ state, $\left|\alpha\right\rangle$, $\left|\beta\right\rangle$ = $\left|g\right\rangle$, $\left|r\right\rangle$ or $\left|e\right\rangle$.
The second part $H_{I}$ represents the coupling between MR and optical cavity, and coupling between optical cavity and atomic ensemble in the rotating-wave approximation, where the energy non-conserving terms $a^{\dagger}\sigma_{eg}^{j}$ and $a\sigma_{ge}^{j}$ have been dropped.
The last part $H_{pump}$ describes the coherent pumping of the optical cavity and atomic ensemble.

In the rotating frame at the laser frequencies $\omega_{p}$ and $\omega_{r}$, the Hamiltonian transforms into a time-independent form,
\begin{eqnarray}
H & = & -\delta'_{c}a^{\dagger}a+\omega_{m}b^{\dagger}b-\Delta_{g}\Sigma_{j=1}^{N}\sigma_{ee}^{j}-\Delta_{gr}\Sigma_{j=1}^{N}\sigma_{rr}^{j}+\lambda_{0}a^{\dagger}a\left(b^{\dagger}+b\right)\nonumber \\
 &  & +g_{0}\Sigma_{j=1}^{N}\left(a\sigma_{eg}^{j}+a^{\dagger}\sigma_{ge}^{j}\right)+\Omega_{p}\left(a^{\dagger}+a\right)+\Omega_{r}\Sigma_{j=1}^{N}\left(\sigma_{er}^{j}+\sigma_{re}^{j}\right),
\label{eq:Hamiltonian_in_rotating_frame}
\end{eqnarray}
where $\ensuremath{\delta'_{c}=\omega_{p}-\omega_{c}}$ is the frequency detuning between optical mode and the cavity drive.

For the atomic ensemble, it is useful to denote collective bosonic operators \cite{Aspelmeyer2014, Chen2015},
$E=\ensuremath{\Sigma_{j=1}^{N}\sigma_{ge}^{\left(j\right)}/\sqrt{N}}$,
$R=\ensuremath{\Sigma_{j=1}^{N}\sigma_{gr}^{\left(j\right)}/\sqrt{N}}$,
with $\ensuremath{\left[E,E^{\dagger}\right]=\left[R,R^{\dagger}\right]=1}$.
Then Hamiltonian can be simplified to
\begin{eqnarray}
H & = & -\delta'_{c}a^{\dagger}a+\omega_{m}b^{\dagger}b-\Delta_{g}E^{\dagger}E-\Delta_{gr}R^{\dagger}R+\lambda_{0}a^{\dagger}a\left(b^{\dagger}+b\right)\nonumber \\
 &  & +g_{N}\left(aE^{\dagger}+a^{\dagger}E\right)+\Omega_{p}\left(a^{\dagger}+a\right)+\Omega_{r}\left(E^{\dagger}R+ER^{\dagger}\right),
\label{eq:Hamiltonian_after_bosonization_approximation}
\end{eqnarray}
where $g_{N}=g_{0}\sqrt{N}$ is the collective atom-cavity coupling rate.

Under the assumption of strong pumping ($\left|\bar{a}\right|\gg 1$), we can rewrite each operator $O$ as the sum of its steady-state solution $\bar{O}$ and a small fluctuation $\delta O$, i.e., $O=\bar{O}+\delta O$ ($O=a,b,E,R$).
Thus, the linearized Hamiltonian can be given as (Note that, hereafter we denote $\delta O$ as $O$ for simplicity):
\begin{eqnarray}
H_{L} & = & -\delta_{c}a^{\dagger}a+\omega_{m}b^{\dagger}b-\Delta_{g}E^{\dagger}E-\Delta_{gr}R^{\dagger}R+\lambda\left(a^{\dagger}+a\right)\left(b^{\dagger}+b\right)\nonumber \\
 &  & +g_{N}\left(aE^{\dagger}+a^{\dagger}E\right)+\Omega_{r}\left(E^{\dagger}R+ER^{\dagger}\right),
\label{eq:Hamiltonian_after_linearization_approximation}
\end{eqnarray}
where $\ensuremath{\delta_{c}=\delta'_{c}-\lambda_{0}\left(\bar{b}+\bar{b}^{*}\right)}$ is the effective detuning modified by the optomechanical coupling, $\ensuremath{\lambda=\lambda_{0}\bar{a}}$ is the optomechanical coupling rate enhanced by the cavity field.
And the steady-state solutions of operators are
\begin{eqnarray}
\bar{a} & = & \Omega_{p}\left[\Omega_{r}^{2}-\Delta_{gr}\left(\Delta_{g}+i\gamma\right)\right]/D,\nonumber \\
\bar{b} & = & -\lambda_{0}\left|\bar{a}\right|^{2}/\omega_{m},\nonumber \\
\bar{E} & = & -g_{N}\Omega_{p}\Delta_{gr}/D,\nonumber \\
\bar{R} & = & -g_{N}\Omega_{p}\Omega_{r}/D,
\label{eq:steady-state_solutions_of_operators}
\end{eqnarray}
where $D=\Omega_{r}^{2}\left(\delta_{c}+i\kappa\right)+\Delta_{gr}\left[g_{N}^{2}-\left(\delta_{c}+i\kappa\right)\left(\Delta_{g}+i\gamma\right)\right]$.

\subsection{Langevin equations and master equation}

To calculate cooling dynamics, we can write the Langevin equations of the system operators, based on the linearized Hamiltonian $H_{L}$,
\begin{eqnarray}
\dot{a} & = & \left(i\delta_{c}-\kappa\right)a-ig_{N}E-i\lambda\left(b^{\dagger}+b\right)+\sqrt{2\kappa}a_{in},\nonumber \\
\dot{b} & = & \left(-i\omega_{m}-\gamma_{m}\right)b-i\lambda\left(a^{\dagger}+a\right)+\sqrt{2\gamma_{m}}b_{in},\nonumber \\
\dot{E} & = & \left(i\Delta_{g}-\gamma\right)E-i\left(g_{N}a+\Omega_{r}R\right)+\sqrt{2\gamma}E_{in},\nonumber \\
\dot{R} & = & i\Delta_{gr}R-i\Omega_{r}E,
\label{eq:Langevin_equations}
\end{eqnarray}
where $2\kappa$, $2\gamma_{m}$ and $2\gamma$ represent the energy decay rate of the optical mode $a$, mechanical mode $b$ and the atomic collective mode $E$, respectively.
$a_{in}$, $b_{in}$ and $E_{in}$ are the corresponding noise inputs from the bath, and satisfy
\begin{eqnarray}
\left\langle a_{in}\left(t\right)a_{in}^{\dagger}\left(t'\right)\right\rangle  & = & \left\langle E_{in}\left(t\right)E_{in}^{\dagger}\left(t'\right)\right\rangle =\delta\left(t-t'\right),\nonumber \\
\left\langle a_{in}^{\dagger}\left(t\right)a_{in}\left(t'\right)\right\rangle  & = & \left\langle E_{in}^{\dagger}\left(t\right)E_{in}\left(t'\right)\right\rangle =0,\nonumber \\
\left\langle b_{in}\left(t\right)b_{in}^{\dagger}\left(t'\right)\right\rangle  & = & \left(n_{th}+1\right)\delta\left(t-t'\right),\nonumber \\
\left\langle b_{in}^{\dagger}\left(t\right)b_{in}\left(t'\right)\right\rangle  & = & n_{th}\delta\left(t-t'\right).
\end{eqnarray}
Here $n_{th}=1/\left[e^{\hbar\omega_{m}/k_{B}T}-1\right]$ is the thermal phonon number at mechanical bath temperature $T$, where $\hbar\approx1.055\times10^{-34}{\rm J\cdot s}$ is the reduced Planck constant, and $k_{B}\approx1.381\times10^{-23}{\rm J\cdot K^{-1}}$ is the Boltzmann constant.

We can also give the master equation of the system,
\begin{eqnarray}
\fl\dot{\rho}   =   -i\left[H_{L},\rho\right]+2\gamma_{m}\left\{ \left(n_{th}+1\right)\mathcal{L}\left[b\right]\rho+n_{th}\mathcal{L}\left[b^{\dagger}\right]\rho\right\}
+2\kappa\mathcal{L}\left[a\right]\rho+2\gamma\mathcal{L}\left[E\right]\rho,
\label{eq:master_equation}
\end{eqnarray}
where $\mathcal{L}\left[O\right]\rho=\frac{1}{2}\left(2O\rho O^{\dagger}-O^{\dagger}O\rho-\rho O^{\dagger}O\right)$ is the Lindblad superoperator, $H_{L}$ in the first term is the linearized Hamiltonian, the second, third and last terms describe the mechanical, optical and atomic decay.

Analogous to \cite{Yi2014,Zhang2014b}, our scheme also focuses on the Lamb-Dicke regime, wherein the optomechanical coupling rate $\lambda$ is sufficiently weaker than the mechanical frequency $\omega_m$, for example $\lambda=0.02\omega_m$.
Then based on the Langevin equations and master equation, theoretical and numerical cooling results can be obtained, through the quantum noise approach \cite{Clerk2010} and covariance matrix approach \cite{Wilson-Rae2008}, respectively.

\section{Theoretical results: Quantum noise approach  \label{sec:Analytical_Results}}

Quantum noise approach permits us to solve the model and gain insight into the cooling results.
We can derive the noise spectrum of optical force, which is related to the heating/cooling coefficient.
By analyzing the upper bound of noise spectrum, we expect to get the optimal parameter conditions for practical experiments to control directly a small heating coefficient, which is far less than its supremum and a large cooling coefficient, which is very close to its supremum. As a result, a small final phonon number can be achieved.

\subsection{Noise spectrum and the upper bound}

The optical force noise spectrum is defined as
$S_{FF}\left(\omega\right)=\int_{-\infty}^{+\infty}dt e^{i\omega t}
\left\langle F\left(t\right)F\left(0\right)\right\rangle$,
and can be calculated in the frequency domain by \cite{Liu2013}
\begin{equation}
S_{FF}\left(\omega\right)=\int_{-\infty}^{+\infty}d\omega'F^{\dagger}\left(\omega\right)F\left(\omega'\right).
\label{eq:calculation_of_noise_spectrum}
\end{equation}
Here $F$ represents the optical radiation force, $F=-\frac{\partial H_{L}}{\partial x}$ \cite{Aspelmeyer2014},
where $x=x_{ZPF}\left(b+b^{\dagger}\right)$, $x_{ZPF}=\sqrt{\hbar/\left(2m_{eff}\omega_{m}\right)}$ and $m_{eff}$ are the displacement, zero-point fluctuation and effective mass of the MR, respectively, thus
\begin{equation}
F\left(\omega\right)=-\lambda\left[a\left(\omega\right)+a^{\dagger}\left(\omega\right)\right]/x_{ZPF},
\label{eq:optical_radiation_force}
\end{equation}
where $a\left(\omega\right)$ can be solved by the transformation of Langevin equations in the frequency domain,
\begin{eqnarray}
\fl -i\omega a\left(\omega\right) & = & \left(i\delta_{c}-\kappa\right)a\left(\omega\right)-ig_{N}E\left(\omega\right)
-i\lambda\left[\left(b^{\dagger}\right)\left(\omega\right)+b\left(\omega\right)\right]+\sqrt{2\kappa}a_{in}\left(\omega\right),\nonumber \\
\fl -i\omega b\left(\omega\right) & = & \left(-i\omega_{m}-\gamma_m\right)b\left(\omega\right)
-i\lambda\left[\left(a^{\dagger}\right)\left(\omega\right)+a\left(\omega\right)\right]+\sqrt{2\gamma_m}b_{in}\left(\omega\right),\nonumber \\
\fl -i\omega E\left(\omega\right) & = & \left(i\Delta_{g}-\gamma\right)E\left(\omega\right)
-i\left[g_{N}a\left(\omega\right)+\Omega_{r}R\left(\omega\right)\right]+\sqrt{2\gamma}E_{in}\left(\omega\right),\nonumber \\
\fl -i\omega R\left(\omega\right) & = & i\Delta_{gr}R\left(\omega\right)-i\Omega_{r}E\left(\omega\right).
\label{eq:Langevin_equations_in_frequency_domain}
\end{eqnarray}
Due to the assumption of weak coupling, the optomechanical coupling terms can be neglected.
Then, the solution of $a\left(\omega\right)$ is expressed as
\begin{equation}
a\left(\omega\right)=-\frac{1}{\chi}\left[\sqrt{2\kappa}\chi_{1}\left(\omega\right)a_{in}\left(\omega\right)+i\sqrt{2\gamma}g_{N}E_{in}\left(\omega\right)\right],
\label{eq:aw}
\end{equation}
where
\begin{eqnarray}
\chi\left(\omega\right) & = & \chi_{1}\left(\omega\right)\chi_{2}\left(\omega\right)+g_{N}^{2},\nonumber \\
\chi_{1}\left(\omega\right) & = & i\left[\left(\omega+\Delta_{g}\right)-\Omega_{r}^{2}/\left(\omega+\Delta_{gr}\right)\right]-\gamma,\nonumber \\
\chi_{2}\left(\omega\right) & = & i\left(\omega+\delta_{c}\right)-\kappa.
\label{eq:chi_chi1_chi2}
\end{eqnarray}

By plugging the expressions of $F\left(\omega\right)$ (\ref{eq:optical_radiation_force}) and $a\left(\omega\right)$ (\ref{eq:aw}) back into (\ref{eq:calculation_of_noise_spectrum}), we can obtain the explicit expression of $S_{FF}\left(\omega\right)$,
\begin{equation}
S_{FF}\left(\omega\right)=\frac{\left|\lambda\right|^{2}}{x_{ZPF}^{2}}\frac{1}{\left|\chi\left(\omega\right)\right|^{2}}\left[2\kappa\left|\chi_{1}\left(\omega\right)\right|^{2}+2\gamma g_{N}^{2}\right].
\label{eq:expression_of_noise_spectrum}
\end{equation}
To gain a upper bound of $S_{FF}\left(\omega\right)$ for arbitrary $\omega$, we denote $M\left(\omega\right)$ as
\begin{equation}
M\left(\omega\right)=\frac{\kappa\left|\chi_{1}\left(\omega\right)\right|^{2}}{\gamma g_{N}^{2}},
\label{eq:M}
\end{equation}
then (\ref{eq:expression_of_noise_spectrum}) can be transformed into
\begin{eqnarray}
\fl  S_{FF}\left(\omega\right) & =  \frac{\left|\lambda\right|^{2}}{x_{ZPF}^{2}}\frac{2\kappa\left|\chi_{1}\left(\omega\right)\right|^{2}}{\left|\chi\left(\omega\right)\right|^{2}}\left(1+\frac{1}{M}\right)\nonumber \\
  \fl & =    \frac{2\left|\lambda\right|^{2}\kappa}{x_{ZPF}^{2}}\left(1+\frac{1}{M}\right)\left\{\kappa^{2}\left(1+\frac{1}{M}\right)^{2}
 +\left[\omega+\delta_{c}-\frac{g_{N}^{2}{\rm Im}\left[\chi_{1}\left(\omega\right)\right]}{\left|\chi_{1}\left(\omega\right)\right|^{2}}\right]^{2}\right\}^{-1},
\label{eq:Noise_spectrum_in_M_form}
\end{eqnarray}
where ${\rm Im}\left[\cdot\right]$ represents a function picking the imaginary part,
${\rm Im}\left[\chi_{1}\left(\omega\right)\right]=\omega+\Delta_{g}-\Omega_{r}^{2}/\left(\omega+\Delta_{gr}\right)$.
Thus,
\begin{equation}
S_{FF}\left(\omega\right)\le\frac{2\left|\lambda\right|^{2}}{\kappa\left(1+\frac{1}{M\left(\omega\right)}\right)x_{ZPF}^{2}}\equiv S_{FF}^{Up}\left(\omega\right),
\label{eq:supremum_of_noise_specturm}
\end{equation}
where $S_{FF}^{Up}\left(\omega\right)$ is just the expected upper bound, and
$S_{FF}\left(\omega\right)=S_{FF}^{Up}\left(\omega\right)$ if and only if
\begin{equation}
\omega+\delta_{c}-\frac{g_{N}^{2}{\rm Im}\left[\chi_{1}\left(\omega\right)\right]}{\left|\chi_{1}\left(\omega\right)\right|^{2}}=0.
\label{eq:condition_of_reaching_supremum_of_noise_specturm}
\end{equation}

\subsection{Heating/cooling coefficient and mean phonon number}
Based on the Fermi golden rule and noise spectrum, the heating/cooling coefficient $A_{\pm}$ reads \cite{Marquardt2007}

\begin{equation}
A_{\pm}=S_{FF}\left(\mp\omega_{m}\right)x_{ZPF}^{2}.
\label{eq:cooling_and_heating_rate}
\end{equation}
Then, the time evolution of mean phonon number $\left\langle n\right\rangle $ can be given as \cite{Yi2014,Zhang2014b}

\begin{equation}
\left\langle n\right\rangle =\left(n_{th}-n_{ss}\right)e^{-Wt}+n_{ss},
\label{eq:final_and_evolution_of_n}
\end{equation}
where

\begin{eqnarray}
W & = & 2\gamma_m+A_{-}-A_{+},
\label{eq:cooling_rate}\\
n_{ss} & = & \frac{A_{+}+2\gamma_{m}n_{th}}{A_{-}-A_{+}+2\gamma_{m}}.
\label{eq:steady_state_phonon_number}
\end{eqnarray}
Here $n_{th}$ is the thermal phonon number due to the mechanical bath, $W$ is the cooling rate and also called effective mechanical energy damping rate, and $n_{ss}$ is the steady-state phonon number, since $\mathop {\lim }\limits_{t \to \infty }  \left\langle n\right\rangle =n_{ss}$ when $W>0$.

The final phonon number in theoretical can be calculated from (\ref{eq:steady_state_phonon_number}), as long as parameters are given.
While, how to choose parameters to achieve a small phonon number is still an important problem.
For this, we find that the upper bound of noise spectrum can give a satisfying answer, with the following three optimal parameter conditions, which can make $A_+$ far less than its supremum while $A_-$ very close to its supremum.

\subsection{Optimal parameter conditions}
\begin{figure}[t!]
\centering
\includegraphics[width=8cm]{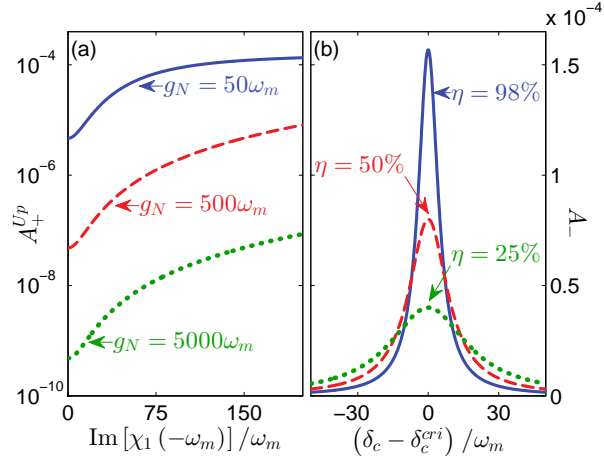}
\protect\caption{(a) The upper bound $A_{+}^{Up}$ of heating coefficient
as a function of ${\rm Im}\left[\chi_{1}\left(-\omega_{m}\right)\right]$ with three different $g_{N}$.
The blue solid, red dashed, and green dotted curves correspond to $g_{N}=50\omega_{m}$, $500\omega_{m}$, and $5000\omega_{m}$.
(b) The cooling coefficient $A_{-}$ as a function of $\delta_{c}-\delta_{c}^{cri}$ with three different $\eta$,
where $\delta_{c}^{cri}$ denotes the solution to (\ref{eq:optimal_cooling_condition_large_cooling_rate}) for $\delta_{c}$,
while $\Delta_{g}$ and $\Delta_{gr}$ satisfy (\ref{eq:optimal_cooling_condition_small_heating_rate}) and $A_{-}=\eta A_{-}^{Sup}$.
The blue solid, red dashed, and green dotted curves correspond to $\eta=98\%$, $50\%$, and $25\%$.
Here $\kappa=5\omega_{m}$, $\gamma=15\omega_{m},$ $\lambda=0.02\omega_{m}$ are used in both panel (a) and (b).
In addition, $g_{N}=5000\omega_{m}$ and $\Omega_{r}=60\omega_{m}$ are used as well in panel (b).
\label{fig:AzAf}}
\end{figure}

On one hand, to get a small heating coefficient $A_+$ directly, based on (\ref{eq:supremum_of_noise_specturm}) we can find
\begin{eqnarray}
A_{+} & = & S_{FF}\left(-\omega_{m}\right)x_{ZPF}^{2}\nonumber \\
 & \le & S_{FF}^{Up}\left(-\omega_{m}\right)x_{ZPF}^{2}=\frac{2\left|\lambda\right|^{2}}{\kappa}\frac{1}{1+M\left(-\omega_m\right)}\equiv A_{+}^{Up},
\label{eq:upper_bound_of_heating_rate}
\end{eqnarray}
where $A_{+}^{Up}$ is the upper bound of $A_+$. Then based on (\ref{eq:M}) and the second line of (\ref{eq:chi_chi1_chi2}), we can make $A_{+}^{Up}$ reach its minimum using the condition ${\rm Im}\left[\chi_{1}\left(-\omega_{m}\right)\right]=0$, i.e.,
\begin{equation}
\Delta_{g}-\omega_{m}-\Omega_{r}^{2}/\left(\Delta_{gr}-\omega_{m}\right)=0.
\label{eq:optimal_cooling_condition_small_heating_rate}
\end{equation}
In this situation,
\begin{equation}
A_{+}^{Up}=\frac{{2{{\left| \lambda  \right|}^2}}}{\kappa }\frac{1}{{1 + C}},
\label{eq:upper_bound_heating_rate_with_C}
\end{equation}
where we have introduced the cooperativity $C\equiv g_{N}^{2}/\left(\kappa\gamma\right)$.
Thus, based on (\ref{eq:upper_bound_of_heating_rate}), (\ref{eq:upper_bound_heating_rate_with_C}) and
\begin{eqnarray}
\lim_{C\rightarrow +\infty}A_{+}^{Up}=0\equiv A_{+}^{Inf},
\label{eq:Inf_limit_of_heating_rate}\\
\lim_{C\rightarrow 0}A_{+}^{Up}=\frac{2\left|\lambda\right|^2}{\kappa}\equiv A_{+}^{Sup},
\label{eq:Sup_limit_of_heating_rate}
\end{eqnarray}
with the assumption of strong collective atom-cavity coupling, i.e., $C\gg1$, we can obtain a small heating coefficient, i.e.,
\begin{equation}
A_+\le A_{+}^{Up}\approx A_{+}^{Inf}\ll A_{+}^{Sup},
\end{equation}
where $A_{+}^{Inf}$ and $A_{+}^{Sup}$ are the infimum and supremum of $A_{+}^{Up}$, respectively.
Figure~\ref{fig:AzAf}(a) shows the variation of $A_{+}^{Up}$ with ${\rm Im}\left[\chi_{1}\left(-\omega_{m}\right)\right]$. As expected from (\ref{eq:optimal_cooling_condition_small_heating_rate}) $A_{+}^{Up}$ reaches its minimum when ${\rm Im}\left[\chi_{1}\left(-\omega_{m}\right)\right]=0$, and as $g_N$ (and $C$) grows $A_{+}^{Up}$ becomes smaller.

On the other hand, to get a large cooling coefficient $A_-$ directly, based on (\ref{eq:supremum_of_noise_specturm}) and (\ref{eq:condition_of_reaching_supremum_of_noise_specturm}) we make $S_{FF}\left(\omega_{m}\right)$ reach its upper bound using the condition
\begin{equation}
\omega_{m}+\delta_{c}-\frac{g_{N}^{2}{\rm Im}\left[\chi_{1}\left(\omega_{m}\right)\right]}{{\rm Im}^{2}\left[\chi_{1}\left(\omega_{m}\right)\right]+\gamma^{2}}=0.
\label{eq:optimal_cooling_condition_large_cooling_rate}
\end{equation}
In this situation,
\begin{eqnarray}
A_{-} & = & S_{FF}\left(\omega_{m}\right)x_{ZPF}^{2}\nonumber \\
 & = & S_{FF}^{Up}\left(\omega_{m}\right)x_{ZPF}^{2}=\frac{2\left|\lambda\right|^{2}}{\kappa}\left[1+\frac{1}{M\left(\omega_{m}\right)}\right]^{-1}\equiv A_{-}^{Up}.
\label{eq:value_of_cooling_rate}
\end{eqnarray}
Here $A_{-}^{Up}$ is the upper bound of $A_-$, and
\begin{eqnarray}
\lim_{M\left(\omega_m\right)\rightarrow0}A_{-}^{Up}=0\equiv A_{-}^{Inf},
\label{eq:Inf_limit_of_cooling_rate}\\
\lim_{M\left(\omega_m\right)\rightarrow+\infty}A_{-}^{Up}=\frac{2\left|\lambda\right|^{2}}{\kappa}\equiv A_{-}^{Sup},
\label{eq:Sup_limit_of_cooling_rate}
\end{eqnarray}
where $A_{-}^{Inf}$ and $A_{-}^{Sup}$ are the infimum and supremum of $A_{-}^{Up}$, respectively.
Then we can make $A_{-}^{Up}$ reach close to its supremum $A_{-}^{Sup}$ using the condition
\begin{equation}
\left[1+\frac{1}{M\left(\omega_{m}\right)}\right]^{-1}\ge\eta,
\label{eq:optimal_cooling_condition_inequality_pre}
\end{equation}
where $0<\eta<1$ is a flexible real number.
In this situation,
\begin{equation}
A_{-}^{Up}\ge\eta A_{-}^{Sup}.
\label{eq:optimal_cooling_condition_inequality}
\end{equation}
Thus, from (\ref{eq:value_of_cooling_rate}) and (\ref{eq:optimal_cooling_condition_inequality}) we can obtain a large cooling coefficient, i.e.,  $A_{-}=A_{-}^{Up}\ge \eta A_{-}^{Sup}$.
As shown in figure~\ref{fig:AzAf}(b), $A_{-}$ reaches its maximum when $\delta_c=\delta_c^{cri}$, where $\delta_c^{cri}$ is the solution to the condition (\ref{eq:optimal_cooling_condition_large_cooling_rate}). And as $\eta$ grows the cooling coefficient $A_{-}$ becomes larger, while as a trade-off, the constraint of parameters condition (\ref{eq:optimal_cooling_condition_inequality}) becomes tighter.
Here, the value of $\eta$ can be assigned directly independent of other parameters, for example one can let $\eta=98\%$, while $\Delta_{g}$, $\Delta_{gr}$, and $\delta_{c}$ can be solved by the optimal parameter conditions [(\ref{eq:optimal_cooling_condition_small_heating_rate}),
(\ref{eq:optimal_cooling_condition_large_cooling_rate}) and
(\ref{eq:optimal_cooling_condition_inequality})]
(See \ref{sec:Appendix_A} for details).

\begin{figure}[t!]
\centering
\includegraphics[width=8cm]{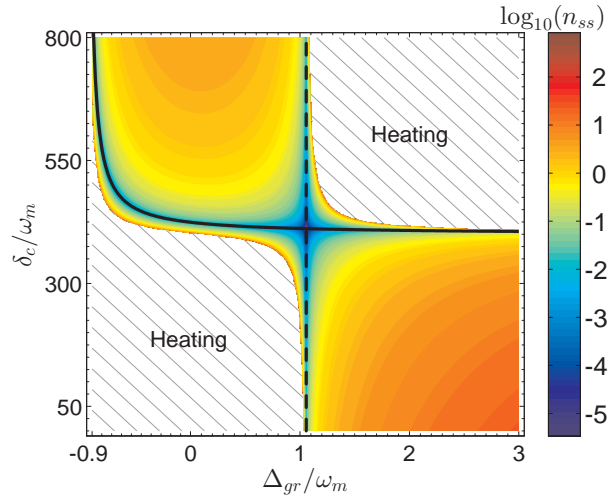}
\protect\caption{Theoretical result of final phonon number $n_{ss}$ versus $\delta_{c}$ and $\Delta_{gr}$,
with $\Delta_{g}$ satisfying $A_{-}=\eta A_{-}^{Sup}$.
The hatched area represents heating the MR.
The black dashed and solid curves correspond to the optimal parameter conditions (\ref{eq:optimal_cooling_condition_small_heating_rate}) and
(\ref{eq:optimal_cooling_condition_large_cooling_rate}), respectively.
Here $\eta=98\%$, $\gamma_{m}=0$,
$\kappa=5\omega_{m}$, $\gamma=15\omega_{m},$ $\lambda=0.02\omega_{m}$, $g_{N}=5000\omega_{m}$ and $\Omega_{r}=60\omega_{m}$.
\label{fig:nss_vd_gr_dc_theoretical}}
\end{figure}

Under these conditions, a small final phonon number in theoretical can be achieved, as shown in figure~\ref{fig:nss_vd_gr_dc_theoretical}.
The solid and dashed curves indicate the condition making the cooling coefficient large and that making the heating coefficient small, respectively. One can see that the phonon number reaches its minimum at the intersection.
Specially, when $\gamma_{m}=0$, from (\ref{eq:steady_state_phonon_number}),
(\ref{eq:upper_bound_of_heating_rate}) and
(\ref{eq:optimal_cooling_condition_inequality}) we obtain
\begin{equation}
n_{ss}=\frac{A_+}{A_{-}-A_{+}}\approx\frac{A_{+}}{A_{-}}<\frac{1}{\eta\left(1+C\right)},
\label{eq:cooling_limit_without_r}
\end{equation}
where $C\gg1$ as mentioned above. This shows that the order of cooling limit in our scheme is the same as those in \cite{Yi2014,Zhang2014b}.

\section{Numerical results: Covariance matrix approach \label{sec:Numerical_Results}}

In order to investigate our model more deeply and verify the theoretical approach, similar to \cite{Chen2015, Liu2015a}, we numerically simulate the cooling process using the covariance matrix approach.
From the master equation (\ref{eq:master_equation}), we can derive the equation of motion of the arbitrary second-order moment $O$ by \cite{Gardiner2000},
\begin{equation}
\frac{\partial\left\langle O\right\rangle }{\partial t}={\rm Tr}\left\{ O\frac{\partial\rho}{\partial t}\right\}.
\label{eq:Calculation_of_MotionE_by_MasterE}
\end{equation}
All the involved second-order moments are:
$\left\langle aa\right\rangle $,
$\left\langle a^{\dagger}a\right\rangle $, $\left\langle ba\right\rangle $,
$\left\langle ba^{\dagger}\right\rangle $, $\left\langle bb\right\rangle $,
$\left\langle b^{\dagger}b\right\rangle $, $\left\langle Ea\right\rangle $,
$\left\langle Ea^{\dagger}\right\rangle $, $\left\langle Eb\right\rangle $,
$\left\langle Eb^{\dagger}\right\rangle $, $\left\langle EE\right\rangle $,
$\left\langle E^{\dagger}E\right\rangle $, $\left\langle Ra\right\rangle $,
$\left\langle Ra^{\dagger}\right\rangle $, $\left\langle Rb\right\rangle $,
$\left\langle Rb^{\dagger}\right\rangle $, $\left\langle RE\right\rangle $,
$\left\langle RE^{\dagger}\right\rangle $, $\left\langle RR\right\rangle $,
$\left\langle R^{\dagger}R\right\rangle $,
where $\left\langle \cdot \right\rangle $ denotes the average value.
Then the system of all motion equations (See \ref{sec:Appendix_B} for details) can be solved numerically through the four-order Runge-Kutta algorithm.

\begin{figure}[b!]
\centering
\includegraphics[width=8cm]{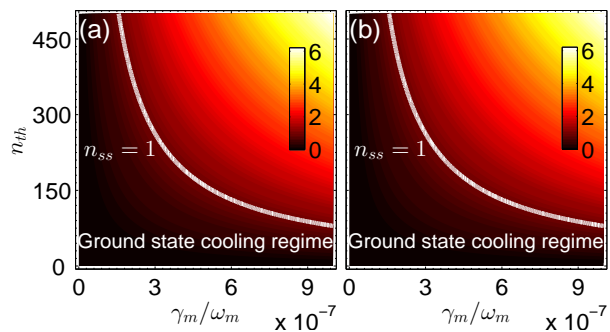}
\protect\caption{(a) Numerical and (b) theoretical results of
final phonon number $n{}_{ss}$ as a function of
the mechanical damping rate $\gamma_{m}$ and
the thermal phonon number $n_{th}$.
The white solid curves correspond to $n_{ss}=1$,
marking the boundary between ground state cooling regime and the opposite.
Here $\eta=98\%$,
$\kappa=5\omega_{m}$, $\gamma=15\omega_{m}$,
$\lambda=0.02\omega_{m}$, $g_{N}=5000\omega_{m}$, $\Omega_{r}=60\omega_{m}$,
and the detunings $\Delta_{g}$, $\Delta_{gr}$ and $\delta_{c}$ satisfying $A_{-}=\eta A_{-}^{Sup}$, (\ref{eq:optimal_cooling_condition_small_heating_rate}) and (\ref{eq:optimal_cooling_condition_large_cooling_rate}).
\label{fig:nss_rm_nth_theoretical_numerical}}
\end{figure}

The above theoretical results obtained by quantum noise approach rely on the assumption $\lambda\ll\omega_m$.
To testify the reasonability of this assumption, we numerically solve the equations of motion (\ref{eq:all_motion_equations}) for the time evolution of mean phonon number, and plot the dependence of final phonon number $n_{ss}$ on the thermal phonon number $n_{th}$ and the mechanical damping rate $\gamma_m$ in figure~\ref {fig:nss_rm_nth_theoretical_numerical}(a), which shows a good agreement with the theoretical result [Figure~ \ref{fig:nss_rm_nth_theoretical_numerical}(b)] from the expression of $n_{ss}$ (\ref{eq:steady_state_phonon_number}).
In addition, as we can see, the smaller $n_{th}$ and $\gamma_m$, the smaller $n_{ss}$, and vice versa.
Strictly speaking, from (\ref{eq:steady_state_phonon_number}) we can obtain ground state cooling, i.e., $n_{ss}<1$ requires
\begin{equation}
n_{th}<\frac{A_{-}-2A_{+}}{2\gamma_{m}}+1,
\label{eq:requirement_of_nth}
\end{equation}
or
\begin{equation}
\gamma_{m}<\frac{A_{-}-2A_{+}}{2\left(n_{th}-1\right)}.
\label{eq:requirement_of_Gamma_m}
\end{equation}
Then one can find that, the upper bound of tolerable $n_{th}$ or $\gamma_m$ for ground state cooling becomes larger when the optomechanical coupling rate $\lambda$ increases, due to $A_{-},A_{+}\propto\left|\lambda\right|^{2}$ based on (\ref{eq:expression_of_noise_spectrum}) and (\ref{eq:cooling_and_heating_rate}).
For the current feasible parameters in the published experiments, the mechanical quality factor $Q_{m}=\omega_{m}/\gamma_{m}>5\times10^{6}$ and temperature of the bath $T\approx 20{\rm mK}$ \cite{Moser2014,Yuan2015,Norte2016}, with the corresponding $\gamma_{m}<2\times10^{-7}\omega_{m}$, $n_{th}\approx 416$, ground state cooling of the MR is achievable under our optimal parameter conditions.

\begin{figure}[b!]
\centering
\includegraphics[width=8cm]{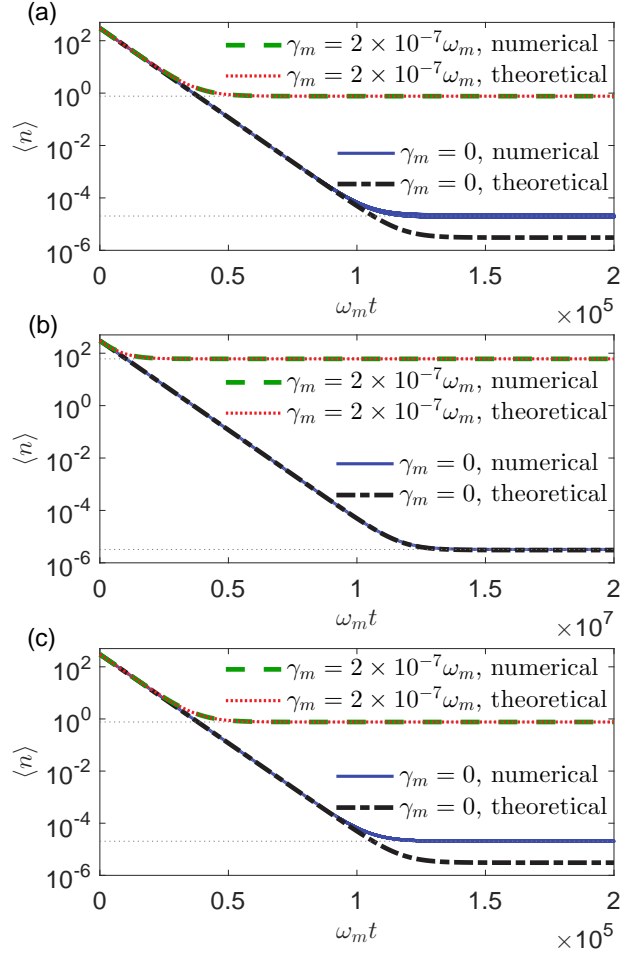}
\protect\caption{Time evolution of the mean phonon number, both in numerical and theoretical.
The green dashed, red dotted, blue solid, and black dash-dotted curves correspond to the results with
$\gamma_{m}=2\times10^{-7}\omega_{m}$ in numerical,
$\gamma_{m}=2\times10^{-7}\omega_{m}$ in theoretical,
$\gamma_{m}=0$ in numerical and
$\gamma_{m}=0$ in theoretical, respectively.
The horizontal gray dashed lines are the corresponding steady-state solutions of the motion equations (\ref{eq:all_motion_equations}).
In panel (a) $\kappa=5\omega_m$, $g_N=5\times10^3\omega_m$ and $\lambda=0.02\omega_m$.
In panel (b) $\kappa=500\omega_m$, $g_N=5\times10^4\omega_m$ and $\lambda=0.02\omega_m$.
In panel (c) $\kappa=500\omega_m$, $g_N=5\times10^4\omega_m$ and $\lambda=0.2\omega_m$.
The other common parameters are $\eta=98\%$, $n_{th}=300$,
$\gamma=15\omega_{m}$, $\Omega_{r}=60\omega_{m}$.
The detunings $\Delta_{g}$, $\Delta_{gr}$ and $\delta_{c}$ satisfy $A_{-}=\eta A_{-}^{Sup}$, (\ref{eq:optimal_cooling_condition_small_heating_rate}) and (\ref{eq:optimal_cooling_condition_large_cooling_rate}).
\label{fig:Time_evolution_of_n}}
\end{figure}

It is interesting to study the effects of mechanical damping $\gamma_m$, cavity decay $\kappa$ and cavity-enhanced optomechanical coupling $\lambda$ on the cooling performance.
Figure \ref{fig:Time_evolution_of_n}(a), \ref{fig:Time_evolution_of_n}(b) and \ref{fig:Time_evolution_of_n}(c) plot the numerical results of the mean phonon number $\left\langle n\right\rangle =\left\langle b^{\dagger}b\right\rangle $ as a function of time for $\gamma_m=0$ and $\gamma_m=2\times 10^{-7}\omega_m$, as well as the corresponding theoretical results, with parameters (a) $\kappa=5\omega_m$, $\lambda=0.02\omega_m$, (b) $\kappa=500\omega_m$, $\lambda=0.02\omega_m$ and (c) $\kappa=500\omega_m$, $\lambda=0.2\omega_m$, respectively. Note that, the corresponding values of $g_N$ make $C=g_{N}^{2}/\left(\kappa\gamma\right)$ remains unchanged for different $\kappa$.

From figure \ref{fig:Time_evolution_of_n}(a), we can see that when $\gamma_m=0$, the steady state of $\left\langle n\right\rangle$ achieves the order of $10^{-5}$. When $\gamma_m=2\times 10^{-7}\omega_m$, due to heating of the bath, the steady state phonon number increases close to $10^0$, while as a trade-off, the time required to achieve the steady state is reduced. This is because the cooling rates $W$ ($=  2\gamma_m+A_{-}-A_{+}$) for $\gamma_m=0$ and $\gamma_m=2\times 10^{-7}\omega_m$ are nearly the same under the optimal parameter conditions.

Compared figure \ref{fig:Time_evolution_of_n}(b) to \ref{fig:Time_evolution_of_n}(a), we can see that for $\kappa=500\omega_m$, the time spent cooling to the steady state is close to 100 times as that for $\kappa=5\omega_m$.  When $\gamma_m=0$ the final phonon number $n_{ss}$ stays unchanged, while when $\gamma_m=2\times10^{-7}\omega_m$, $n_{ss}$ increases.
Actually, under the assumption $C\gg1$, from (\ref{eq:upper_bound_of_heating_rate}) and (\ref{eq:optimal_cooling_condition_inequality}), we find that
\begin{equation}
A_+\approx \frac{{2{{\left| \lambda  \right|}^2}}}{\kappa }\frac{1}{{1 + C}},\qquad A_-\approx\frac{{2{{\left| \lambda  \right|}^2}}}{\kappa }.
\label{eq:approx_heating_cooling_coefficient}
\end{equation}
Then when $\kappa$ changes to $\kappa'$, the heating/cooling coefficient changes to $A'_{\pm} \approx {A_ \pm }\kappa/\kappa'$, which results in the new cooling rate and new final phonon number
\begin{eqnarray}
W' \approx W\kappa/\kappa',
\label{eq:new_cooling_rate}\\
n{'_{ss}} \approx \frac{{{A_ + } + 2{{\gamma'}_m}{n_{th}}}}{{{A_ - } - {A_ + } + 2{{\gamma'}_m}}},
\label{eq:new_final_phonon_number}
\end{eqnarray}
where $\gamma'_m=\gamma_m\kappa'/\kappa$.
Thus, for $\kappa'=500\omega_m$ and $\kappa=5\omega_m$, after some analysis one can find that figure \ref{fig:Time_evolution_of_n}(b) is consistent with the theoretical expectation. Moreover, here comes to a conclusion that the larger $\kappa$, the larger required mechanical quality $Q_m$ ($=\omega_m/\gamma_m$), and vice versa.

Compared figure \ref{fig:Time_evolution_of_n}(c) to \ref{fig:Time_evolution_of_n}(a), we can see that for $\kappa=500\omega_m$ and $\lambda=0.2\omega_m$, the cooling dynamics is almost the same as that for $\kappa=5\omega_m$ and $\lambda=0.02\omega_m$. Actually, this can be understood easily from (\ref{eq:approx_heating_cooling_coefficient}). Thus, we can obtain that for a big cavity decay $\kappa$, it is helpful for ground state cooling to increase $\lambda$ properly.
In addition, as shown in figure \ref{fig:Time_evolution_of_n}(a), \ref{fig:Time_evolution_of_n}(b) and \ref{fig:Time_evolution_of_n}(c), the numerical curves are in good agreement with the corresponding theoretical ones. The reason for the slight difference is that in the theoretical calculation the multi-phonon process is not considered.

\section{Discussion \label{sec:Discussion}}

In this section, we will give some comparisons between our scheme and other related works \cite{Genes2011,Bariani2014}.

Compared with \cite{Genes2011}, there are several similarities between the system configurations of the two schemes: both the schemes use a three-level atomic ensemble and an optical cavity to cool the MR. However, the physical mechanism to achieve cooling in our scheme differs significantly from \cite{Genes2011}.
Generally speaking, in \cite{Genes2011} C. Genes {\it et al} tune the cavity resonance to the Anti-Stokes through $\delta_c=\Delta_{gr}=-\omega_m$, consequently the Anti-Stokes sideband is enhanced, which yields a large cooling coefficient. Moreover, they get an effective sharpening of the Lorentzian cavity response through the assumption $g_N\gg\Omega_r$, consequently the off-resonant Stokes sideband is further suppressed, which yields a small heating coefficient.
While in our scheme, based on the quantum noise approach, on one hand we first make the upper bound of the heating coefficient $A_{+}$ reach its minimum through the condition (\ref{eq:optimal_cooling_condition_small_heating_rate}), then make this minimum come close to its infimum $A_{+}^{Inf}$ (\ref{eq:Inf_limit_of_heating_rate}) and far away from its supremum $A_{+}^{Sup}$ (\ref{eq:Sup_limit_of_heating_rate}) through the assumption $C\gg 1$, and thus obtain a small heating coefficient. On the other hand, we first make the cooling coefficient $A_{-}$ reach its upper bound through the condition (\ref{eq:optimal_cooling_condition_large_cooling_rate}), then make this upper bound reach close to its supremum $A_{-}^{Sup}$ (\ref{eq:Sup_limit_of_cooling_rate}) through the condition (\ref{eq:optimal_cooling_condition_inequality}), and thus obtain a large cooling coefficient.
The main advantages of our scheme over \cite{Genes2011} are discussed in detail as follows.

\begin{figure}[b!]
\centering
\includegraphics[width=8cm]{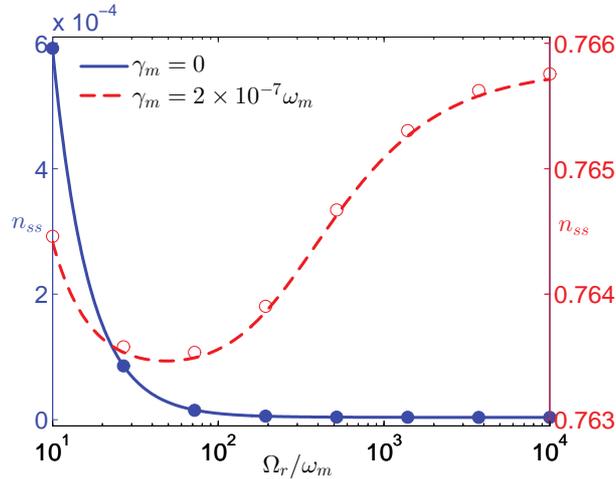}
\protect\caption{Numerical results of the final phonon number $n_{ss}$ as a function of the atomic drive strength $\Omega_{r}$,
with the detunings $\Delta_{g}$, $\Delta_{gr}$ and $\delta_{c}$ satisfying $A_{-}=\eta A_{-}^{Sup}$, (\ref{eq:optimal_cooling_condition_small_heating_rate}) and (\ref{eq:optimal_cooling_condition_large_cooling_rate}).
The blue solid curve and blue full circles correspond to the steady-state solutions and the time evolution solutions of $\gamma_{m}=0$.
The red dashed curve and red open circles correspond to those solutions of $\gamma_{m}=2\times10^{-7}\omega_{m}$.
Here, $\eta=98\%$, $n_{th}=300$,
$\kappa=5\omega_{m}$, $\gamma=15\omega_{m}$,
$\lambda=0.02\omega_{m}$, $g_{N}=5000\omega_{m}$ and  $10\omega_{m}\le\Omega_{r}\le10^{4}\omega_{m}$.
\label{fig:nss_omr_rm_0_rm_neq_0}}
\end{figure}

Firstly, our scheme assumes $g_N^2\gg \kappa\gamma$, i.e., the cooperativity $C\gg 1$, rather than $g_N\gg\Omega_r$ in \cite{Genes2011}, although our scheme also works in the strong collective atom-cavity coupling regime.
Actually, based on our theory and under the optimal parameter conditions [(\ref{eq:optimal_cooling_condition_small_heating_rate}),
(\ref{eq:optimal_cooling_condition_large_cooling_rate}) and
(\ref{eq:optimal_cooling_condition_inequality})],
\begin{equation}
\fl A_+=S_{FF}\left(-\omega_m\right)x_{ZPF}^2=\frac{2\left|\lambda\right|^{2}\kappa\left(1+C\right)}
{\kappa^{2}\left(1+C\right)^{2}+\left(\delta_{c}-\omega_{m}\right)^{2}}, \qquad
A_-\ge\eta\frac{2\left|\lambda\right|^2}{\kappa},
\label{eq:cooling_heating_rate_under_optimal}
\end{equation}
where $\delta_c$ is independent of $\Omega_r$ due to
\begin{eqnarray}
\eta
& \mathop\Rightarrow\limits ^{\left(\ref{eq:value_of_cooling_rate}\right)} &
M\left(\omega_{m}\right)=\frac{\eta}{1-\eta}
\mathop\Rightarrow\limits ^{\left(\ref{eq:M}\right)}
{\rm Im}\left[\chi_{1}\left(\omega_{m}\right)\right]=
\sqrt{\frac{\gamma g_N^2}{\kappa}M\left(\omega_{m}\right)-\gamma^2} \nonumber\\
& \mathop\Rightarrow\limits ^{\left(\ref{eq:optimal_cooling_condition_large_cooling_rate}\right)} &
\delta_c=\frac{g_{N}^{2}{\rm Im}\left[\chi_{1}\left(\omega_{m}\right)\right]}{{\rm Im}^{2}\left[\chi_{1}\left(\omega_{m}\right)\right]+\gamma^{2}}-\omega_{m}.
\label{eq:dc_under_optimal_conditions}
\end{eqnarray}
Consequently, from the expression of final phonon number (\ref{eq:steady_state_phonon_number}), we can find that theoretical result of $n_{ss}$ is independent of $\Omega_r$.
Figure \ref{fig:nss_omr_rm_0_rm_neq_0} plots the numerical results of $n_{ss}$ as a function of $\Omega_r$, where $\Delta_{g}$, $\Delta_{gr}$ and $\delta_{c}$ satisfy the optimal parameter conditions.
One can see that, for both $\gamma_m=0$ and $\gamma_m=2\times 10^{-7}\omega_m$, when $\Omega_r$ varies from $10\omega_m$ to $10^4\omega_m$, $n_{ss}$ only changes on the order of $10^{-4}$, which can be neglected.
Thus both the theoretical and numerical results verify that in our scheme, the value of $\Omega_r$ can be chosen freely according to the practical experimental conditions, regardless of the limitation $\Omega_r\ll g_N$.

\begin{figure}[b!]
\centering
\includegraphics[width=8cm]{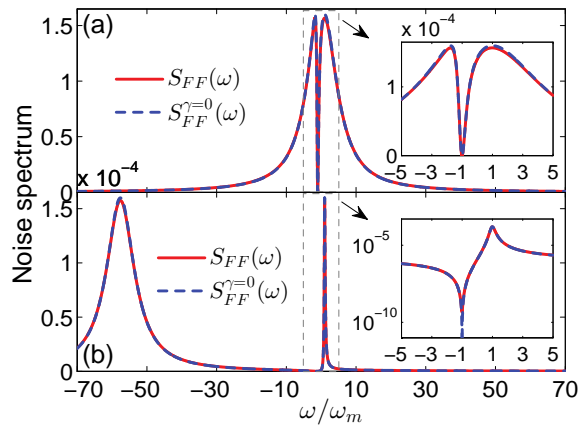}
\protect\caption{Noise spectrum $S_{FF}\left(\omega\right)$ and its
approximation $S_{FF}^{\gamma=0}\left(\omega\right)$ as a function of $\omega$  for (a) $\Omega_{r}=15\omega_{m}$ and
(b) $\Omega_{r}=150\omega_{m}$, with the detunings $\Delta_{g}$, $\Delta_{gr}$ and $\delta_{c}$ satisfying $A_{-}=\eta A_{-}^{Sup}$, (\ref{eq:optimal_cooling_condition_small_heating_rate}) and (\ref{eq:optimal_cooling_condition_large_cooling_rate}).
The red solid and blue dashed curves represent $S_{FF}\left(\omega\right)$
and $S_{FF}^{\gamma=0}\left(\omega\right)$, respectively.
The insets are the closeup view of the EIT and Fano regions.
Here $\eta=98\%$,
$\kappa=5\omega_{m}$, $\gamma=15\omega_{m}$,
$\lambda=0.02\omega_{m}$, $g_{N}=5000\omega_{m}$ and $\Omega_{r}=60\omega_{m}$.
\label{fig:Sffw_and_Sffw_r_equals_0}}
\end{figure}

The direct reason for the absence of $\Omega_r\ll g_N$ is that, the condition (\ref{eq:optimal_cooling_condition_small_heating_rate}) resulting in a small heating coefficient can be seen as a constraint of $\Delta_g$ and $\Delta_{gr}$, without any limitation on $\Omega_r$.
While, a more fundamental reason can be extracted from the noise spectrum. In \cite{Genes2011}, the cavity response is a Lorentzian lineshape with the center at $\omega=\omega_m$. However, from figure~\ref{fig:Sffw_and_Sffw_r_equals_0} we can see that the noise spectrum in our scheme similar to \cite{Chen2015,Liu2015a,Guo2014} can be non-Lorentzian lineshapes, for example, symmetric electromagnetically induced transparency (EIT) or asymmetric Fano lineshape. A peak is located at $\omega=\omega_m$, meanwhile, a dip also appears at $\omega=-\omega_m$, which directly suppresses the Stokes sideband and thus leads to a better cooling performance.
Actually, the lineshape of the noise spectrum is determined by the locations of extreme points, which can be approximately calculated from the maximum condition
\begin{equation}
\omega+\delta_{c}-\frac{g_{N}^{2}}{\left(\omega+\Delta_{g}\right)-\Omega_{r}^{2}/\left(\omega+\Delta_{gr}\right)}=0,
\label{eq:Maximum_Condition_sffw_r0}
\end{equation}
and the minimum condition
\begin{equation}
\left(\omega+\Delta_{g}\right)-\Omega_{r}^{2}/\left(\omega+\Delta_{gr}\right)=0,
\label{eq:Minimum_Condition_sffw_r0}
\end{equation}
of the approximation of $S_{FF}\left(\omega\right)$, which is denoted as
\begin{eqnarray}
S_{FF}^{\gamma=0}\left(\omega\right)
 & \equiv & \frac{2\left|\lambda\right|^{2}\kappa}{x_{ZPF}^{2}}\left\{ \left[\omega+\delta_{c}-\frac{g_{N}^{2}}{\omega+\Delta_{g}-\frac{\Omega_{r}^{2}}{\omega+\Delta_{gr}}}\right]^{2}+\kappa^{2}\right\} ^{-1}.
\label{eq:Noise_Spectrum_r_equals_0}
\end{eqnarray}
Here the deduction of $S_{FF}^{\gamma=0}\left(\omega\right)$ relies on (\ref{eq:Noise_spectrum_in_M_form}) and $M\gg1$, $\gamma\ll{\rm Im}\left[\chi_{1}\left(\omega\right)\right]$, which are valid under our optimal parameter conditions apart from a small region around $\omega=-\omega_m$. Figure~\ref{fig:Sffw_and_Sffw_r_equals_0} shows that $S_{FF}^{\gamma=0}\left(\omega\right)$ agrees well with $S_{FF}\left(\omega\right)$, which verifies our understanding of the lineshape of noise spectrum.

Secondly, different from \cite{Genes2011}, which requires that the number of atoms is much larger than the number of intracavity steady-state photons, i.e., $N\gg\left|\bar{a}\right|^{2}$, due to $g_N=g_0\sqrt{N}\gg\Omega_r$ and the weak atomic excitation assumption $\Omega_r\gg g_{0}\left|\bar{a}\right|$ in \cite{Genes2011}, in our scheme this limitation can be removed by increase the pump strength $\Omega_r$ of atomic ensemble. The reason is that, strictly speaking the steady-state atomic excitation should be much smaller than the number of atoms $N$, i.e.,
\begin{equation}
\left|E\right|^{2}+\left|R\right|^{2}\ll N,
\label{eq:standard_weak_atomic_excition_condition}
\end{equation}
where $\left|E\right|^2$ and $\left|R\right|^2$ are the steady-state excitation of atomic $\left|e\right\rangle$ and $\left|r\right\rangle$ energy levels, respectively. Based on (\ref{eq:steady-state_solutions_of_operators}), we can find
\begin{equation}
\frac{\left|\bar{a}\right|}{\left|\bar{E}\right|}\approx\frac{\Omega_{r}^{2}}{2g_{N}\omega_{m}},\qquad \frac{\left|\bar{a}\right|}{\left|\bar{R}\right|}\approx\frac{\Omega_{r}}{2g_{N}},
\label{eq:relation_steady_state_variables}
\end{equation}
thus from (\ref{eq:standard_weak_atomic_excition_condition}) and (\ref{eq:relation_steady_state_variables}), a condition between $N$ and $\left|\bar{a}\right|$ can be given as
\begin{equation}
N\gg\left|\bar{a}\right|^{2}\left\{ \frac{4g_{N}^{2}}{\Omega_{r}^{2}}\left[1+\frac{\omega_{m}^{2}}{\Omega_{r}^{2}}\right]\right\}.
\label{eq:relation_N_a_bar_our_scheme}
\end{equation}
Thus, in \cite{Genes2011} due to $g_N\gg\Omega_r$, $N\gg\left|\bar{a}\right|^{2}$. However, in our scheme by increasing $\Omega_r$, the required number of atoms $N$ can be reduced, which is valid because $n_{ss}$ is basically independent of $\Omega_r$, as shown in figure~\ref{fig:nss_omr_rm_0_rm_neq_0}. In addition, for a fixed $N$ by increasing $\Omega_r$, the tolerable $\left|\bar{a}\right|^{2}$ can be increased, which benefits the promotion of cooling rate $W$ (\ref{eq:cooling_rate}) due to $S_{FF}\left(\omega\right)\propto\left|\bar{a}\right|^{2}$ (\ref{eq:Noise_spectrum_in_M_form}).
Note that, the number of atoms in our scheme also satisfies
\begin{equation}
N\gg\left(\frac{\kappa\gamma}{g_{0}}\right)^{2},
\label{eq:assumption_N_our_scheme}
\end{equation}
due to the assumption of cooperativity $C\gg1$, which is familiar in many atom-optomechanical cooling works \cite{Zeng2015,Yi2014,Zhang2014b}.

Finally, compared with \cite{Genes2011}, the ranges of parameters $\delta_{c}$ and $\Delta_{gr}$ are extended in our scheme.
The reason is that we have introduced a flexible parameter $\eta$ ($0<\eta<1$) in our condition (\ref{eq:optimal_cooling_condition_inequality}).
When $\eta\rightarrow 1$, $\delta_{c}$ and $\Delta_{gr}$ in our scheme are equivalent to those in \cite{Genes2011}, i.e., $\delta_c,\Delta_{gr}\rightarrow -\omega_m$, due to
\begin{eqnarray}
\eta\rightarrow1
& \mathop\Rightarrow\limits ^{\left(\ref{eq:optimal_cooling_condition_inequality}\right)} &
A_{-}\rightarrow A_{-}^{Sup}
\mathop\Rightarrow\limits ^{\left(\ref{eq:value_of_cooling_rate}\right)}
M\left(\omega_{m}\right)\rightarrow+\infty\nonumber \\
& \mathop\Rightarrow\limits ^{\left(\ref{eq:M}\right)} &
{\rm Im}\left[\chi_{1}\left(\omega_{m}\right)\right]\rightarrow\infty
\left\{
\begin{array}{l}
\mathop\Rightarrow\limits ^{\left(\ref{eq:optimal_cooling_condition_large_cooling_rate}\right)}
\omega_{m}+\delta_{c}\rightarrow0.
\\
\mathop\Rightarrow\limits ^{\left(\ref{eq:chi_chi1_chi2}\right)}
\omega_{m}+\Delta_{gr}\rightarrow0.
\end{array}
\right.
\end{eqnarray}
While, for other cases of $\eta$, from (\ref{eq:dc_under_optimal_conditions}) we can find that the value of $\delta_c$ and $\Delta_{gr}$ will not be limited at $-\omega_m$.
For example, with $\kappa=5\omega_{m}$, $\gamma=15\omega_{m}$ and $g_{N}=5\times10^3\omega_{m}$, when $\eta=98\%$, $\delta_c\approx411.39$ and when $\eta=99\%$, $\delta_c\approx289.13$.
Note that as a trade-off, the cooling coefficient $A_-$ in our scheme is $\left(1-\eta\right)\times 100$ percent smaller than its supremum.
Thus, in experiments, one can balance the range of parameters and the value of cooling coefficient to choose an appropriate $\eta$.

Compared with \cite{Bariani2014}, both the schemes essentially exploit a three level atomic ensemble to tailor the cavity response or the optical force noise spectrum, and then achieve cooling of the MR. However, the couplings between the atomic ensemble and the optomechanical cavity are different. In our scheme, atomic ensemble is coupled to the optomechanical cavity directly. By contrast, in \cite{Bariani2014}, atomic ensemble indirectly couples to the optomechanical cavity, mediated by an extra optical cavity with a much larger decay. Therefore, due to cooling in our scheme does not rely on the coupling between the two cavities, it is easier to be realized in experiment. In addition, a broader parameter space is achieved for ground state cooling in our scheme. As we can see from figure~\ref{fig:Time_evolution_of_n}(c), the steady-state phonon number $n_{ss}$ for $\kappa=500\omega_m$ can be smaller than 1, almost the same as that for $\kappa=5\omega_m$ in figure~\ref{fig:Time_evolution_of_n}(a). While results in \cite{Bariani2014} show that in the highly unresolved sideband regime, the MR is cooled to a final phonon number still far away from the ground state. This difference is probably resulted from the corresponding adjustment of other parameters ($g_N$ and $\lambda$) based on the quantum noise approach in our scheme for large $\kappa$. Note that, in \cite{Bariani2014} a cooling method based on the motional states of atomic ensemble is also proposed, which is shown to have a better cooling performance than that based on the atomic energy level structure, while as mentioned there, the tunability is diminished to some extent.

\section{Conclusion \label{sec:Conclusion}}

In conclusion, we have investigated the cooling of MR in a hybrid optomechanical system with a three-level atomic ensemble. In the Lamb-Dicke regime, the cooling dynamics are derived by using the quantum noise approach. To achieve the ground state, we find three optimal parameter conditions, under which a large cooling coefficient and a small heating coefficient can be obtained through the calculation of noise spectrum, under the assumption of strong collective atom-cavity coupling. Moreover, through the covariance matrix approach, numerical simulations demonstrate that ground state cooling is feasible experimentally, even in the highly unresolved sideband regime. In addition, compared with the existing MR cooling method with a three-level atomic ensemble \cite{Genes2011}, in our scheme there are almost no limitations on the drive strength of atomic ensemble and the number of atoms, and the tolerable parameters space for ground state cooling are extended. This method may provide a guideline for cooling of MR in the hybrid optomechanical system \cite{Chen2015,Zeng2015,Yi2014,Zhang2014b,Bariani2014} both for theoretical and experimental research.

\section*{Acknowledgments}

We are grateful to the reviewers for constructive suggestions, and we thank R. S. He and J. Lin for stimulating discussions. This work was supported by the National Natural Science Foundation of China (Grant No.~11504430 and No.~61502526), and the National Basic Research Program of China (Grant No.~2013CB338002).

\appendix
\section{
The solutions to optimal parameter conditions
[(\ref{eq:optimal_cooling_condition_small_heating_rate}),
(\ref{eq:optimal_cooling_condition_large_cooling_rate}) and
(\ref{eq:optimal_cooling_condition_inequality})]
for the detunings $\Delta_{g}$, $\Delta_{gr}$, and $\delta_{c}$
\label{sec:Appendix_A}
}

First of all, from the condition (\ref{eq:optimal_cooling_condition_small_heating_rate}),
we can write $\Delta_{gr}$ as an expression of $\Delta_{g}$,
\begin{equation}
\Delta_{gr}=\frac{\Omega_{r}^{2}}{\Delta_{g}-\omega_{m}}+\omega_{m}.
\label{eq:optimal_cooling_condition_small_heating_rate_solution}
\end{equation}
Then, combined with $M\left(\omega\right)$ (\ref{eq:M}),
we obtain
\begin{eqnarray}
M\left(\omega_{m}\right)
& = & \frac{\kappa}{\gamma g_{N}^{2}}\left\{ {\rm Im^{2}}\left[\chi_{1}\left(\omega_{m}\right)\right]+\gamma^{2}\right\} \nonumber \\
& = & \frac{\kappa}{\gamma g_{N}^{2}}\left\{ \left[\left(\Delta_{g}+\omega_{m}\right)-\frac{\Omega_{r}^{2}}{\frac{\Omega_{r}^{2}}{\Delta_{g}-\omega_{m}}+2\omega_{m}}\right]^{2}+\gamma^{2}\right\} \nonumber \\
& = & \frac{\kappa}{\gamma g_{N}^{2}}\left\{ \left[\frac{2\omega_{m}\left(\Delta_{g}-\omega_{m}\right)^{2}}{\Omega_{r}^{2}+2\omega_{m}\left(\Delta_{g}-\omega_{m}\right)}+2\omega_{m}\right]^{2}+\gamma^{2}\right\} .
\label{eq:M_in_Appendix}
\end{eqnarray}
And because the inequality condition
(\ref{eq:optimal_cooling_condition_inequality}) or
(\ref{eq:optimal_cooling_condition_inequality_pre})
is equivalent to
\begin{equation}
M\left(\omega_{m}\right)\ge\frac{\eta}{1-\eta}.
\label{eq:optimal_cooling_condition_inequality_in_Appendix}
\end{equation}
Therefore, by plugging (\ref{eq:M_in_Appendix}) into (\ref{eq:optimal_cooling_condition_inequality_in_Appendix}),
we obtain
\begin{equation}
\left[\frac{2\omega_{m}\left(\Delta_{g}-\omega_{m}\right)^{2}}{\Omega_{r}^{2}+2\omega_{m}\left(\Delta_{g}-\omega_{m}\right)}+2\omega_{m}\right]^{2}\ge\frac{\gamma g_{N}^{2}}{\kappa}\frac{\eta}{1-\eta}-\gamma^{2}\equiv\eta'.
\end{equation}
Then, $\Delta_{g}$ satisfies
\begin{equation}
\sigma_{1}\Delta_{g}^{2}+\sigma_{2}\Delta_{g}+\sigma_{3}\ge0,
\label{eq:optimal_cooling_condition_inequality_equiv_1}
\end{equation}
or
\begin{equation}
\Delta_{g}<\omega_{m}-\frac{\Omega_{r}^{2}}{2\omega_{m}}\thinspace\thinspace{\rm and}\thinspace\thinspace\sigma'_{1}\Delta_{g}^{2}+\sigma'_{2}\Delta_{g}+\sigma'_{3}\ge0,
\label{eq:optimal_cooling_condition_inequality_equiv_2}
\end{equation}
where
\begin{eqnarray}
\sigma_{1} & = & \frac{2\omega_{m}}{\sqrt{\eta'}-2\omega_{m}},\nonumber \\
\sigma'_{1} & = & \frac{2\omega_{m}}{\sqrt{\eta'}+2\omega_{m}},\nonumber \\
\sigma_{2} & = & \frac{-2\sqrt{\eta'}\omega_{m}}{\sqrt{\eta'}-2\omega_{m}},\nonumber \\
\sigma'_{2} & = & \frac{2\sqrt{\eta'}\omega_{m}}{\sqrt{\eta'}+2\omega_{m}},\nonumber \\
\sigma_{3} & = & \frac{2\left(\sqrt{\eta'}-\omega_{m}\right)\omega_{m}^{2}}{\sqrt{\eta'}-2\omega_{m}}-\Omega_{r}^{2},\nonumber \\
\sigma'_{3} & = & \frac{-2\left(\sqrt{\eta'}+\omega_{m}\right)\omega_{m}^{2}}{\sqrt{\eta'}+2\omega_{m}}+\Omega_{r}^{2}.
\end{eqnarray}
Thus, as long as the parameters $\eta$, $\kappa$, $\gamma$, $g_N$ and $\Omega_r$ are given, the range of $\Delta_g$ to $A_{-}\ge\eta A_{-}^{Sup}$ (\ref{eq:optimal_cooling_condition_inequality})
can be solved by the two inequalities above.
Without loss of generality, we take a critical solution for $\Delta_{g}$, which makes the inequality hold, i.e., $A_{-}=\eta A_{-}^{Sup}$.
Then $\Delta_{gr}$ and $\delta_{c}$ can be solved by the other two optimal
parameter conditions
(\ref{eq:optimal_cooling_condition_small_heating_rate})
and (\ref{eq:optimal_cooling_condition_large_cooling_rate}).

\section{The motion equations of second-order moments
\label{sec:Appendix_B}}

Based on the master equation (\ref{eq:master_equation}) and the formula to calculate arbitrary second-order moment (\ref{eq:Calculation_of_MotionE_by_MasterE}), all the equations of motion can be derived as follows:
\begin{eqnarray}
\fl \frac{{\rm d}}{{\rm d}t}\left\langle aa\right\rangle  = 2\left(i\delta_{c}-\kappa\right)\left\langle aa\right\rangle -2ig_{N}\left\langle Ea\right\rangle -2i\lambda\left(\left\langle ba^{\dagger}\right\rangle ^{*}+\left\langle ba\right\rangle \right),\nonumber \\
\fl \frac{{\rm d}}{{\rm d}t}\left\langle a^{\dagger}a\right\rangle  = -2\kappa\left\langle a^{\dagger}a\right\rangle -ig_{N}\left(\left\langle Ea^{\dagger}\right\rangle -\left\langle Ea^{\dagger}\right\rangle ^{*}\right)-i\lambda\left(\left\langle ba\right\rangle ^{*}-\left\langle ba\right\rangle +\left\langle ba^{\dagger}\right\rangle -\left\langle ba^{\dagger}\right\rangle ^{*}\right),\nonumber \\
\fl \frac{{\rm d}}{{\rm d}t}\left\langle ba\right\rangle  = \left[i\left(\delta_{c}-\omega_{m}\right)-\kappa-\gamma_{m}\right]\left\langle ba\right\rangle -ig_{N}\left\langle Eb\right\rangle -i\lambda\left(\left\langle aa\right\rangle +\left\langle a^{\dagger}a\right\rangle +\left\langle bb\right\rangle +\left\langle b^{\dagger}b\right\rangle +1\right),\nonumber \\
\fl \frac{{\rm d}}{{\rm d}t}\left\langle ba^{\dagger}\right\rangle  = -\left[i\left(\delta_{c}+\omega_{m}\right)+\kappa+\gamma_{m}\right]ba^{\dagger}+ig_{N}\left\langle Eb^{\dagger}\right\rangle ^{*}\nonumber \\
\fl \qquad \qquad  -i\lambda\left(\left\langle aa\right\rangle ^{*}+\left\langle a^{\dagger}a\right\rangle -\left\langle bb\right\rangle -\left\langle b^{\dagger}b\right\rangle \right),\nonumber \\
\fl \frac{{\rm d}}{{\rm d}t}\left\langle bb\right\rangle  = -2\left(i\omega_{m}+\gamma_{m}\right)\left\langle bb\right\rangle -2i\lambda\left(\left\langle ba\right\rangle +\left\langle ba^{\dagger}\right\rangle \right),\nonumber \\
\fl \frac{{\rm d}}{{\rm d}t}\left\langle b^{\dagger}b\right\rangle  = -2\gamma_{m}\left\langle b^{\dagger}b\right\rangle +2\gamma_{m}n_{th}-i\lambda\left(\left\langle ba\right\rangle ^{*}-\left\langle ba\right\rangle +\left\langle ba^{\dagger}\right\rangle ^{*}-\left\langle ba^{\dagger}\right\rangle \right),\nonumber \\
\fl \frac{{\rm d}}{{\rm d}t}\left\langle Ea\right\rangle  = \left[i\left(\Delta_{g}+\delta_{c}\right)-\kappa-\gamma\right]\left\langle Ea\right\rangle -ig_{N}\left\langle aa\right\rangle -ig_{N}\left\langle EE\right\rangle -i\Omega_{r}\left\langle Ra\right\rangle \nonumber \\
\fl  \qquad \qquad -i\lambda\left(\left\langle Eb^{\dagger}\right\rangle +\left\langle Eb\right\rangle \right),\nonumber \\
\fl \frac{{\rm d}}{{\rm d}t}\left\langle Ea^{\dagger}\right\rangle  = \left[i\left(\Delta_{g}-\delta_{c}\right)-\kappa-\gamma\right]\left\langle Ea^{\dagger}\right\rangle -ig_{N}\left\langle a^{\dagger}a\right\rangle +ig_{N}\left\langle E^{\dagger}E\right\rangle -i\Omega_{r}\left\langle Ra^{\dagger}\right\rangle \nonumber \\
\fl \qquad \qquad  +i\lambda\left(\left\langle Eb^{\dagger}\right\rangle +\left\langle Eb\right\rangle \right),\nonumber \\
\fl \frac{{\rm d}}{{\rm d}t}\left\langle Eb\right\rangle  = \left[i\left(\Delta_{g}-\omega_{m}\right)-\gamma-\gamma_{m}\right]\left\langle Eb\right\rangle -ig_{N}\left\langle ba\right\rangle -i\Omega_{r}\left\langle Rb\right\rangle -i\lambda\left(\left\langle Ea^{\dagger}\right\rangle +\left\langle Ea\right\rangle \right),\nonumber \\
\fl \frac{{\rm d}}{{\rm d}t}\left\langle Eb^{\dagger}\right\rangle  = \left[i\left(\Delta_{g}+\omega_{m}\right)-\gamma-\gamma_{m}\right]\left\langle Eb^{\dagger}\right\rangle -ig_{N}\left\langle ba^{\dagger}\right\rangle ^{*}-i\Omega_{r}\left\langle Rb^{\dagger}\right\rangle \nonumber \\
\fl \qquad \qquad  +i\lambda\left(\left\langle Ea^{\dagger}\right\rangle +\left\langle Ea\right\rangle \right),\nonumber \\
\fl \frac{{\rm d}}{{\rm d}t}\left\langle EE\right\rangle  = 2\left(i\Delta_{g}-\gamma\right)\left\langle EE\right\rangle -2ig_{N}\left\langle Ea\right\rangle -2i\Omega_{r}\left\langle RE\right\rangle ,\nonumber \\
\fl \frac{{\rm d}}{{\rm d}t}\left\langle E^{\dagger}E\right\rangle  = -2\gamma\left\langle E^{\dagger}E\right\rangle -ig_{N}\left(\left\langle Ea^{\dagger}\right\rangle ^{*}-\left\langle Ea^{\dagger}\right\rangle \right)-i\Omega_{r}\left(\left\langle RE^{\dagger}\right\rangle -\left\langle RE^{\dagger}\right\rangle ^{*}\right),\nonumber \\
\fl \frac{{\rm d}}{{\rm d}t}\left\langle Ra\right\rangle  = \left[i\left(\Delta_{gr}+\delta_{c}\right)-\kappa\right]\left\langle Ra\right\rangle -ig_{N}\left\langle RE\right\rangle -i\Omega_{r}\left\langle Ea\right\rangle -i\lambda\left(\left\langle Rb^{\dagger}\right\rangle +\left\langle Rb\right\rangle \right),\nonumber \\
\fl \frac{{\rm d}}{{\rm d}t}\left\langle Ra^{\dagger}\right\rangle  = \left[i\left(\Delta_{gr}-\delta_{c}\right)-\kappa\right]\left\langle Ra^{\dagger}\right\rangle +ig_{N}\left\langle RE^{\dagger}\right\rangle -i\Omega_{r}\left\langle Ea^{\dagger}\right\rangle +i\lambda\left(\left\langle Rb^{\dagger}\right\rangle +\left\langle Rb\right\rangle \right),\nonumber \\
\fl \frac{{\rm d}}{{\rm d}t}\left\langle Rb\right\rangle  = \left[i\left(\Delta_{gr}-\omega_{m}\right)-\gamma_{m}\right]\left\langle Rb\right\rangle -i\Omega_{r}\left\langle Eb\right\rangle -i\lambda\left(\left\langle Ra^{\dagger}\right\rangle +\left\langle Ra\right\rangle \right),\nonumber \\
\fl \frac{{\rm d}}{{\rm d}t}\left\langle Rb^{\dagger}\right\rangle  = \left[i\left(\Delta_{gr}+\omega_{m}\right)-\gamma_{m}\right]\left\langle Rb^{\dagger}\right\rangle -i\Omega_{r}\left\langle Eb^{\dagger}\right\rangle +i\lambda\left(\left\langle Ra^{\dagger}\right\rangle +\left\langle Ra\right\rangle \right),\nonumber \\
\fl \frac{{\rm d}}{{\rm d}t}\left\langle RE\right\rangle  = \left[i\left(\Delta_{gr}+\Delta_{g}\right)-\gamma\right]\left\langle RE\right\rangle -ig_{N}\left\langle Ra\right\rangle -i\Omega_{r}\left\langle RR\right\rangle -i\Omega_{r}\left\langle EE\right\rangle ,\nonumber \\
\fl \frac{{\rm d}}{{\rm d}t}\left\langle RE^{\dagger}\right\rangle  = \left[i\left(\Delta_{gr}-\Delta_{g}\right)-\gamma\right]\left\langle RE^{\dagger}\right\rangle +ig_{N}\left\langle Ra^{\dagger}\right\rangle +i\Omega_{r}\left\langle R^{\dagger}R\right\rangle -i\Omega_{r}\left\langle E^{\dagger}E\right\rangle ,\nonumber \\
\fl \frac{{\rm d}}{{\rm d}t}\left\langle RR\right\rangle  = 2i\Delta_{gr}\left\langle RR\right\rangle -2i\Omega_{r}\left\langle RE\right\rangle ,\nonumber \\
\fl \frac{{\rm d}}{{\rm d}t}\left\langle R^{\dagger}R\right\rangle  = i\Omega_{r}\left(\left\langle RE^{\dagger}\right\rangle -\left\langle RE^{\dagger}\right\rangle ^{*}\right).
\label{eq:all_motion_equations}
\end{eqnarray}

\section*{References}

\end{document}